\documentclass[aps,twocolumn,preprintnumbers]{revtex4}
\usepackage{graphicx} 
\usepackage[version=3]{mhchem}
\usepackage{pifont}
\usepackage{amsmath}
\usepackage{physics}
\usepackage{tabularx}
\usepackage{mathtools}
\usepackage{siunitx}

\newcommand*{\citen}[1]{%
  \begingroup
    \romannumeral-`\x % remove space at the beginning of \setcitestyle
    \setcitestyle{numbers}%
    \cite{#1}%
  \endgroup
}

\begin{document}

\title{Atomistic origins of asymmetric charge-discharge kinetics in off-stoichiometric \ce{LiNiO_2}}
\author{Penghao Xiao}
\email{Penghao.Xiao@dal.ca}
\author{Ning Zhang}
\author{Harold Smith Perez}
\author{Minjoon Park}
\affiliation{Department of Physics and Atmospheric Science, Dalhousie University, Halifax, Nova Scotia, Canada B3H 4R2}

\date{\today}

%\begin{document}

% ---------------------------------------------------------------------------------
\begin{abstract}
\ce{LiNiO_2} shows poor Li transport kinetics at the ends of charge and discharge in the first cycle, which significantly reduces its available capacity in practice. The atomistic origins of these kinetic limits have not been fully understood. Here, we examine Li transport in \ce{LiNiO_2} by first-principles-based kinetic Monte Carlo simulations where both long time scale and large length scale are achieved, enabling direct comparison with experiments. Our results reveal the rate-limiting steps at both ends of the voltage scan and distinguish the differences between charge and discharge at the same Li content. The asymmetric effects of excess Ni in the Li layer (\ce{Ni_{Li}}) are also captured in our unified modelling framework. In the low voltage region, the first cycle capacity loss due to high overpotential at the end of discharge is reproduced without empirical input. While the Li concentration gradient is found responsible for the low overpotential during charge at this state of charge. \ce{Ni_{Li}} increases the overpotential of discharge but not charge because it only impedes Li diffusion in a particular range of Li concentration and does not change the equilibrium voltage profile. The trends from varying the amount of \ce{Ni_{Li}} and temperature agree with experiments. In the high voltage region, charge becomes the slower process. The bottleneck becomes moving a Li from the Li-rich phase (H2) into the Li-poor phase (H3), while the Li hopping barriers in both phases are relatively low. The roles of preexisting nucleation sites and \ce{Ni_{Li}} are discussed. These results provide new atomistic insights of the kinetic hindrances, paving the road to unleash the full potential of high-Ni layered oxide cathodes.   
\end{abstract}

\maketitle 

%=========================================================================
\section{Introduction}
With the recent rise of electric vehicles, the demand for Li-ion batteries with higher energy density and lower cost keeps increasing. This revives the development of layered oxide cathodes with high Ni content in the transition metal layer. These high-Ni oxides present new challenges such as kinetic hindrance and poor cycle stability. \ce{LiNiO_2} has the same issues and is therefore an ideal model system for understanding the electrochemical behavior of high-Ni oxides. Its electrochemical performance can also serve as a benchmark for element substitution and other modifications. Research on the \ce{LiNiO_2} cathode began in the 1990s and has recently seen a resurgence of interest \cite{dahn1990structure, dahn1991rechargeable, ohzuku1993electrochemistry, li2018updating, phattharasupakun2021correlating, kurzhals2021linio2, mock2021atomistic, bautista2023understanding, bianchini2019there}. \ce{LiNiO_2} undergoes four phases during charge based on powder diffraction, namely H1, M, H2, and H3 \cite{li1993situ, li2018updating}. They all share the same O3 oxygen framework but have different Li-vacancy orderings. At the end of charge about 10\% of Li cannot be extracted \cite{bautista2023understanding}, and at the end of discharge another 10-15\% of Li cannot be inserted back under normal galvanostatic conditions \cite{li2018updating, mock2021atomistic}. The capacity lost during the initial discharge cannot be regained in subsequent cycles, and is therefore referred to as the irreversible capacity loss (IRC). What's worse, excess Ni tends to appear in the Li layer (\ce{Ni_{Li}}) in this material, further exacerbating the IRC \cite{rougier1996optimization, delmas1997behavior, bianchi2001electrochemical, phattharasupakun2021correlating}. One of the appealing features of high-Ni oxides is their large capacity falling within the electrolyte voltage window. Nevertheless, the inaccessible capacity due to sluggish kinetics diminishes this benefit. To unlock the extra capacity, it is crucial to develop a deep understanding of the kinetic hindrances.

The Li-vacancy orderings in \ce{LiNiO_2} have been identified by density function theory (DFT)\cite{arroyo2002first, arroyo2003first, mock2021atomistic}, nuclear magnetic resonance (NMR) \cite{chazel2006coupled, li2021new}, and electron diffraction pattern\cite{peres1999lithium, delmas1999lithium}. On the Li transport, Van der Ven and Ceder have shown that the di-vacancy hop has a lower barrier than the single vacancy hop by DFT calculations \cite{van2000lithium, van2001first}. The kinetic hindrance at the end of discharge can be qualitatively explained by a lack of di-vacancies, but the di-vacancy concentrations at different states of charge have not been quantified. The effect of \ce{Ni_{Li}} has been attributed to shrinking the local Li slab distance and blocking Li passing through its nearest neighboring sites \cite{peres1996relationship, kong2018atomic}. However, the overpotential and its positive correlation with \ce{Ni_{Li}}\% in the low voltage region are more profound during discharge than charge. The charge voltage profile barely changes with \ce{Ni_{Li}}\%, but the discharge overpotential is highly sensitive to \ce{Ni_{Li}}\%. This asymmetric/directional behavior cannot be explained by a constant diffusivity associated with a particular phase, suggesting the existence of at least one hidden factor that differs between charge and discharge \cite{XU20222535}. The poor Li transport at the end of charge has also been attributed to the low diffusivity caused by the shrinkage of Li layer in the H3 phase. Nevertheless, phase boundary propagation could equally impede transport in a two-phase region, as exemplified by the case of \ce{LiFePO_4} \cite{mascaro2017measuring, xiao2018kinetic}. Several intertwined factors influence the kinetics at different stages. To pinpoint the bottlenecks at various conditions, a unified atomistic model that treats all these factors on the equal footing is imperative. Also, the model should produce results that can be validated against experimental data, showcasing its predictive capability. Only through this approach can the model lay the foundation for the rational design of next-generation cathodes.

In this work, a multiscale model of this nature is constructed using DFT-informed kinetic Monte Carlo (KMC) and validated experimentally. The model largely reproduces the charge/discharge voltage profiles with asymmetric hysteresis without any empirical parameters. The predicted first cycle IRC, as well as its dependence on \ce{Ni_{Li}}\% and temperature, agrees well with experiments. The energy of the tetrahedral Li is found to be affected by both its nearest octahedral sites and their neighbors. The occupation of the second neighbors strengthens the position of the directly contacting gate Li and increases the diffusion barrier. This explains the deteriorating kinetics as the Li concentration increases at the end of discharge. Ni at the gate position is immune to the second neighbor occupation, which accounts for the earlier rising of the overpotential during discharge in the presence of \ce{Ni_{Li}}. In contrast, the slow kinetics at the end of charge is found to be limited by the H3 phase nucleation and growth instead of bulk diffusion. Preexisting empty sites near defects or surfaces that disturb the Li ordering could accelerate the charge of the last plateau. Our results pave the road for understanding the effects of degradation and element substitution in high-Ni cathode materials.  

%=========================================================================
\section{Methods}
\subsection{Model parameterization}
The density functional theory calculations are performed with VASP \cite{kresse93_R558, kresse94_14251}. Core electrons are described by the projector augmented wave method \cite{kresse99_1758}. Valence electron wavefunctions are expanded by the plane wave basis set with an energy cutoff of 520$\,$eV \cite{kresse96_11169, kresse96_15}. The exchange-correlation energy is accounted under the generalized-gradient approximation with the Perdew-Burke-Ernzerhof (PBE) functional \cite{perdew91, perdew1996generalized}. A Hubbard U correction is applied on 3d orbitals to mitigate the self-interaction error \cite{dudarev1998electron}. An effective U value of 6.0$\,$eV is chosen for Ni \cite{wang2006oxidation, jain2011formation, das2017first}. Some calculations are verified by the strongly constrained and appropriately normed (SCAN) functional \cite{sun2015strongly, sun2016accurate} with and without the long-range van der Waals (vdW) interaction in the rVV10 form \cite{peng2016versatile}. The k-point sampling is done on a $\Gamma$-centered mesh with a density of 20$\,$\AA. 

The PBE+U energies are used to train a cluster expansion model with the CASM code \cite{sanchez1984generalized,van2002automating, puchala2013thermodynamics}. The cutoff radii for pairs, triplets and quadruplets are 6, 6, and 4$\,$\AA. The L1 norm regularization is employed to select a sparse set of non-zero effective cluster interactions (ECIs) \cite{nelson2013compressive}. A regularization strength of 0.03 is selected to balance between overfitting and underfitting. The energy changes along important diffusion paths are given higher weights to ensure the fitting accuracy. For a single atom hopping process, the changes of the correlation vector and total energy are tiny fractions of the absolute values. It is important to emphasize the ECIs that contribute to these changes. The training data distribution and the convex hulls along three representative lines are summarized in Figure S1 in the Supporting Information (SI). 

\subsection{Kinetic simulation}
The rejection-free KMC is implemented \cite{voter2007introduction}, where cations hop between neighboring octahedral and tetrahedral sites \cite{van2000lithium, van2001first, van2001lithium}. Direct hops between neighboring octahedral sites are not included here. For a single vacancy hop between neighboring octahedral sites, both through a tetrahedral site (TSH) and through a O-O dumbbell/edge (ODH) are local minimum energy paths as converged from careful nudged elastic band (NEB) calculations \cite{henkelman00_9978, henkelman00_9901}, and the TSH barrier is marginally lower. %by 0.014$\,$eV. 
There are two equivalent intermediate tetrahedral sites sharing the O-O edge that the ODH path goes through, and the two THS saddle points are off the tetrahedral centers shifting towards that edge. Therefore, the saddle positions and the barrier heights are very close between the TSH and ODH paths. Since the key information is the total hopping rate between the two octahedral sites, only including the TSH paths is enough to capture the majority of the rate. 
When two Ni occupying the two gate sites that are face-sharing with the two tetrahedral sites (Figure S2(e)), the two TSH paths merge with the ODH path. In that case, Li at either tetrahedral site is assigned the saddle point energy. 

The hopping rate, $r$, is evaluated by the Arrhenius equation with the activation barrier, $E_a$, calculated on the fly based on the Br{\o}nsted--Evans--Polanyi (BEP) principle \cite{evans1938inertia, bronsted1928acid, bligaard2004bronsted}. BEP linearly correlates $E_a$ with the energy difference between the two adjacent local minima, $\Delta E$:
\begin{align}
E_a &= E_{a0} + \frac{1}{2} \Delta E \\
r &= A \exp{(\frac{-E_a}{k_B T})} \label{eq:rate}
\end{align}
where $A$, the prefactor, is set to \SI{e13}{\per\second}, a common approximation for elementary processes in solids \cite{li2019adaptive}; $E_{a0}$ is the intrinsic hopping barrier that depends on the ion type; $k_B$ is the Boltzmann constant; $T$ is the temperature. The $E_{a0}$ is 0.15$\,$eV for Li and 0.50$\,$eV for Ni, respectively \cite{xiao2019understanding}. Combining the above, all the hopping rates between neighboring octahedral and tetrahedral sites are automatically determined based on the local environments.

The ensemble is grand canonical for the surface region and canonical for the bulk. For Li adsorption to the electrode surface, Eq.\ref{eq:rate} becomes 
\begin{align}
r &= A \exp{\frac{-E_{as}}{k_B T}} \exp{-\frac{0.5 (E_i-U)}{k_B T}} \label{eq:BV}
\end{align}
where $E_i$ is the Li energy on a surface site; U is the Li energy in the reservoir (anode), which is equal to the negative of voltage v.s. Li metal; $E_{as}$ is the intrinsic barrier for Li attaching/detaching to the electrode surface. Eq.\ref{eq:BV} is equivalent to the Butler–Volmer equation with the symmetry factor $\beta=0.5$ and exchange current $j_0=A \exp{\frac{-E_{as}}{k_B T}}$. The exact value of $E_{as}$ is not critical because the charge transfer is not the rate-limiting step according to electrochemical impedance spectroscopy measurements \cite{li1996electrochemical}. $E_{as}=0.3$$\,$eV is chosen such that the surface process is faster than most canonical hops but not too fast to cause numerical inefficiency. Three consecutive rows along the b axis are open to the Li reservoir, allowing Li creation and annihilation; Li must hop to/from other areas of the supercell during discharge/charge. The supercell size for the KMC simulations is $32\times 32 \times 8$, containing 8192 \ce{LiNiO_2} units. Periodic boundary conditions are applied in all directions.

To simulate the galvanostatic scan, the voltage applied to the surface region is adjusted according to the current feedback. Specifically, the voltage is incremented by a step size of $\pm$ 5$\,$mV when the current dropped below the threshold value. The charging process is switched to constant voltage at 4.4$\,$V. The current is calculated by linearly fitting the slope of the number of Li atoms in the supercell vs. time for every \SI{0.5}{\micro\second}. The threshold current is set to one Li per \SI{}{\micro\second} for the given supercell, which is equivalent to 0.16$\,$A/cm$^2$. The absolute current might seem high compared to experimental values, but the simulation is for one single particle with a thickness of 9.6$\,$nm along the diffusion direction. Compared to a perfect current control, this scheme has possible voltage overshoots when the Li transport is fast because the voltage does not move backward, but the errors are bounded to the voltage step, 5$\,$mV. 

\subsection{Materials preparation in experiment}
The precursors used for the synthesis of \ce{LiNiO_2} were \SI{15}{\micro\meter} \ce{Ni(OH)_2} obtained from CNGR Advanced Materials Co., Ltd. and \ce{LiOH\cdot H_2O} obtained from Nemaska (99\%). \ce{LiNiO_2} positive electrode materials were made by mixing \ce{Ni(OH)_2} and \ce{LiOH\cdot H_2O} in a mortar and pestle for 10 minutes with the Li:Ni ratio of 1.02. The well-mixed precursors were placed in an alumina tube furnace and heated to \SI{480}{\celsius} for \SI{3}{\hour} with a ramping speed of \SI{10}{\celsius/\minute}. Afterward, the samples were taken out of the furnace and re-grounded before heating at \SI{700}{\celsius} for \SI{20}{\hour} with the same ramping speed. The whole heating process was taken place under pure oxygen flow.

Powder X-ray diffraction (XRD) was measured with a Siemens D5000 diffractometer equipped with a diffracted beam monochromator with Cu K$\mathrm{\alpha}$ radiation. The XRD patterns were collected with a scattering angle range between 15$^{\circ}$  and 70$^{\circ}$  with a step size of 0.02$^{\circ}$  and \SI{3}{\second} per step dwell time. Rietveld refinement was performed using the Rietica software with a R-3m space group to quantify the amount of \ce{Ni_{Li}} \cite{phattharasupakun2021correlating}.

Half coin cells were made to measure the first cycle capacity. The positive electrode recipe consists of 92 wt\% active mass, 4 wt\% PVDF (Arkema, Kynar 301F), and 4 wt\% Super S (Timcal). A solid to N-methyl-2-pyrrolidone (NMP, Sigma Aldrich, 99.5\%) ratio of 1:0.8 was used for the slurry mixing. The obtained slurry was cast on aluminum foil with a 0.006$''$ notch bar. The electrodes were dried in a convection oven at \SI{110}{\celsius} for \SI{2}{\hour} before calendaring at 2000 atm pressure. The single-side electrodes were punched into small disks with a diameter of \SI{12.75}{mm} for coin cell assembly. The active material mass loading was around \SI{10}{mg/cm^2}. 2325-type half-coin cells were assembled in an argon-fill glovebox. Two pieces of lithium foil and two separators (Celgard 2300) were used for the cell construction. The electrolyte used in this study was \SI{1.2}{M} \ce{LiPF_6} dissolved in fluoroethylene carbonate (FEC, Shenzhen Capchem) and dimethyl carbonate (DMC, Shenzhen Capchem) with a volume ratio of 1:4. After assembly, the voltage-specific capacity curves were measured between 3.0 and 4.3$\,$V vs. \ce{Li^+/Li} at 20, 30, and \SI{55}{\celsius}. The C-rate used in this study was C/20 with respect to a 1C current of \SI{200}{mA/g}.

%=========================================================================
\section{Results}
A galvanostatic scan is implemented in the simulation to directly compare with experimental voltage profiles without introducing middle-layer quantities, such as diffusivity, to avoid additional assumptions in empirical models used to extract these values. Voltage profiles are simulated for off-stoichiometric \ce{Li_{1-y}Ni_yNiO2} with different amounts of \ce{Ni_{Li}} (y). Li-Ni exchange is not considered because Neutron diffraction shows no Li in the Ni layer \cite{pouillerie2001structural}, and the defects formed during synthesis are from losing \ce{Li2O2} \cite{mccalla2012lithium}.

\begin{figure*}
    \includegraphics[width=0.8\textwidth]{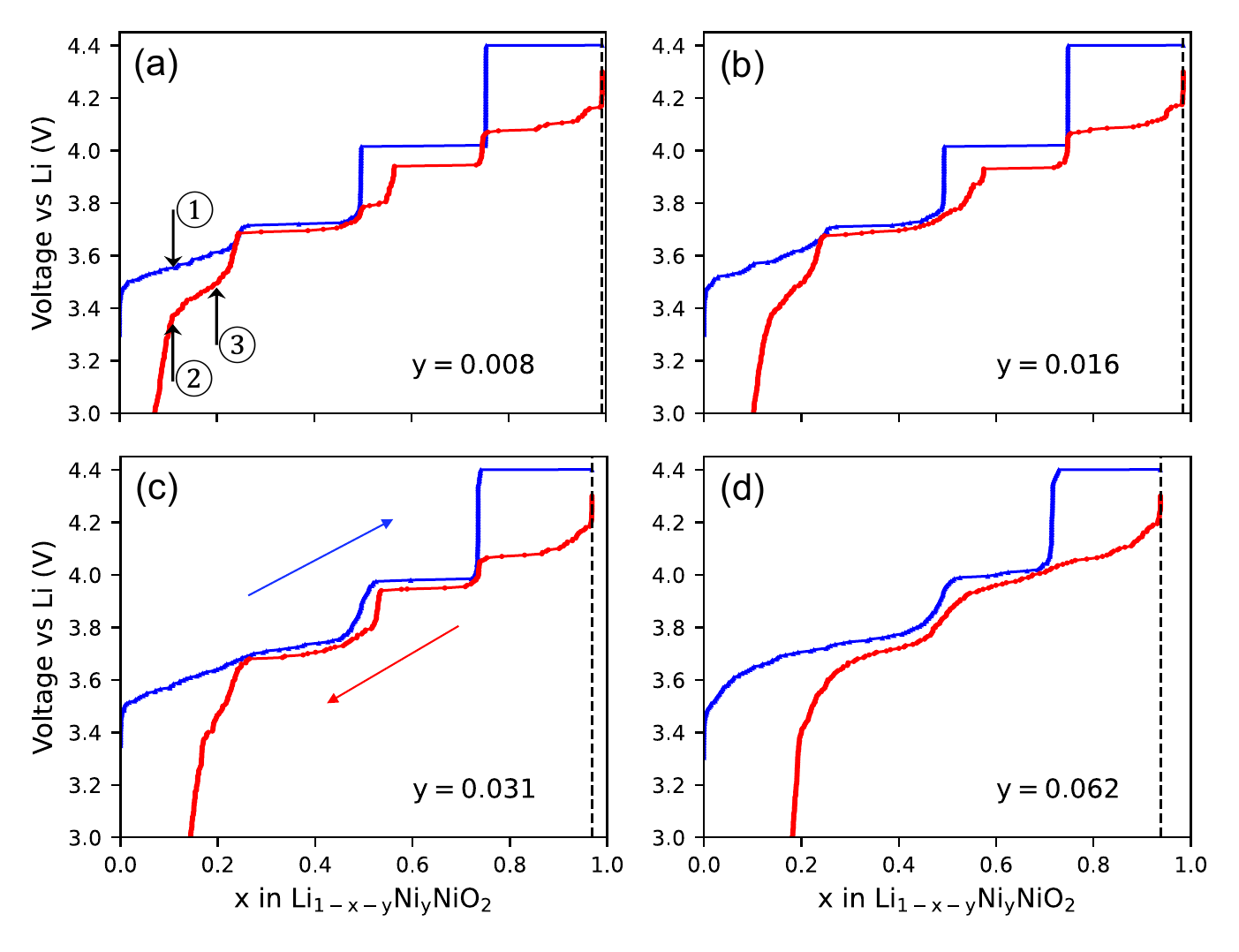}
	\caption{
	Simulated voltage profiles with different \ce{Ni_{Li}} (y) at 320 K.
	\label{fig:voltage_profiles}
	}
\end{figure*}

Figure \ref{fig:voltage_profiles} shows the simulated voltage profiles at four y values. The number of KMC steps for each scan is on the order of $10^9$. The hysteresis between charge and discharge can be clearly seen. For x$<$0.25 (voltage $<$ 3.7$\,$V), the charge curves are close to the equilibrium values (shown in Figure \ref{fig:predictions}), while the discharge curves exhibit increasingly large overpotential as x approaches 0, resulting in the first cycle IRC at 3.0$\,$V voltage cutoff. As y increases from 0.8\% to 6.2\%, the IRC grows from 7\% to 18\%. Despite the surging overpotential in the discharge counterpart, the charge potential increases by less than 0.1$\,$V. Both charge and discharge curves become more sloping as y increases, corresponding to broadened dQ/dV peaks. These trends agree with the experimental observations \cite{delmas1997behavior, phattharasupakun2021correlating, kurzhals2021linio2}. Our model not only captures the IRC from the atomistic level but also reproduces the asymmetry that \ce{Ni_{Li}} affects discharge more than charge. 

\begin{figure*}
    \includegraphics[width=\textwidth]{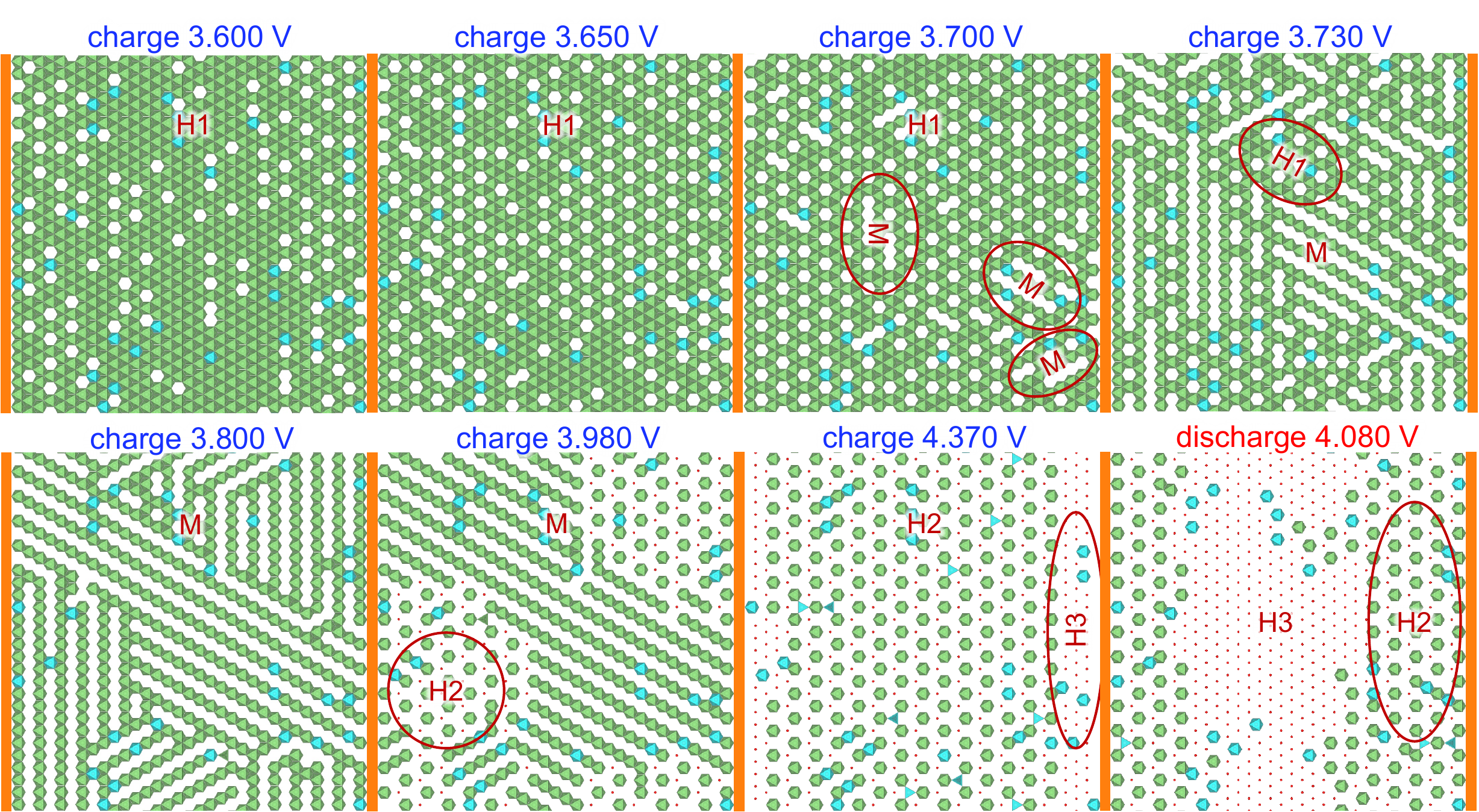}
	\caption{
	Top view of one Li layer during charge and discharge with 3.1\% \ce{Ni_{Li}}. Green polyhedra are Li; cyan polyhedra are Ni; red dots are O. The orange belts mark the open boundaries (surfaces). 
	\label{fig:ordering}
	}
\end{figure*}

\begin{figure*}[hbt!]
    \includegraphics[width=0.8\textwidth]{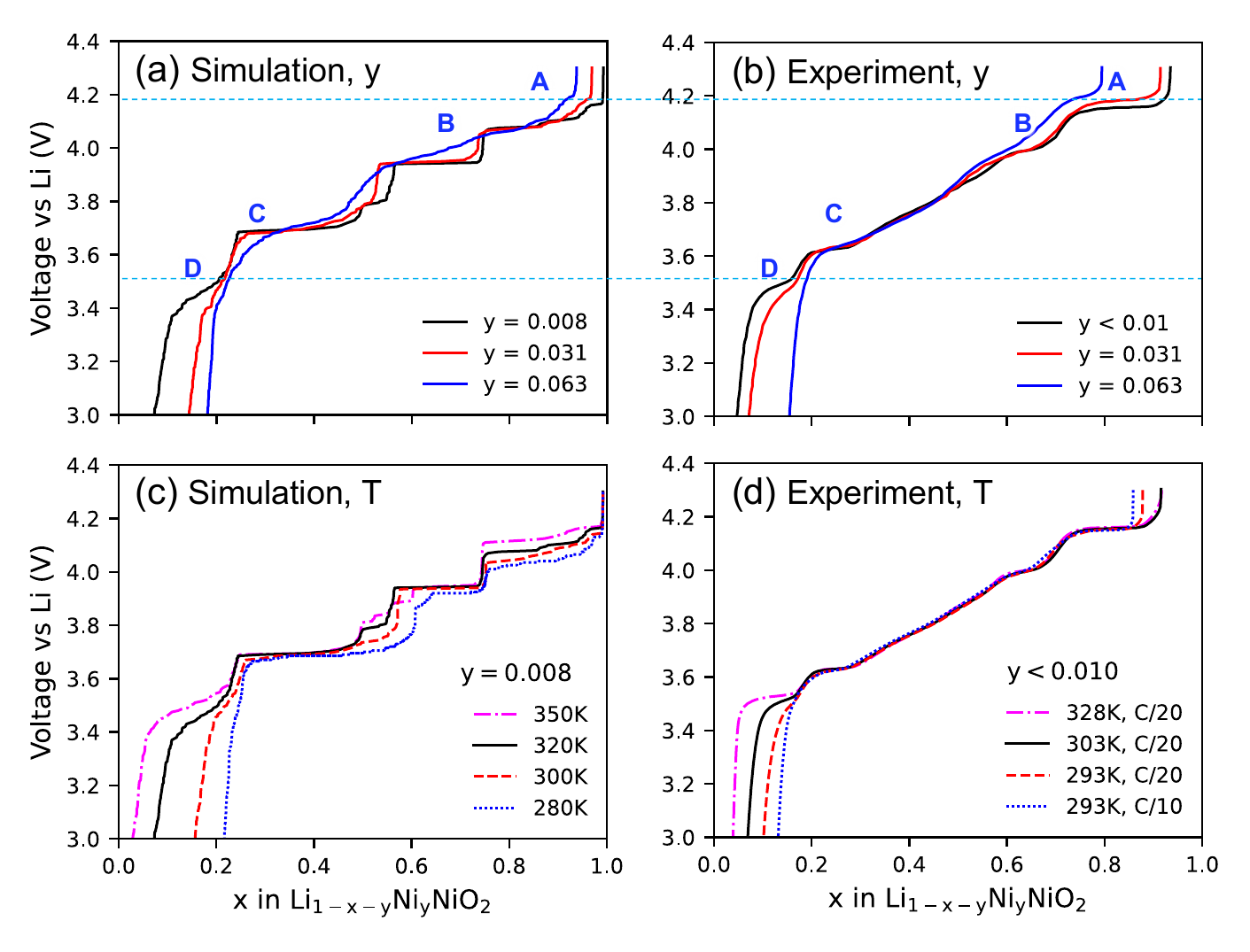}
	\caption{
	Comparison of discharge curves between simulations and  experiments. (a) and (b) compare the effect of \ce{Ni_{Li}} (y). The dashed horizontal lines are a guide to the eye. (c) and (d) compare the effects of temperature and scan rate. 
	\label{fig:discharge compare}
	}
\end{figure*}

Representative Li distribution snapshots are show in Figure \ref{fig:ordering}. The phases are labeled based on the Li-vacancy ordering. The M phase exists from 3.7$\,$V to 4.0$\,$V during charge, in consistent with recent experiments \cite{li2018updating}. One interesting non-equilibrium phenomenon revealed from the simulation is that the M phase forms in three orientations when y$\geq$ 0.031 for the given supercell (particle) size. Indeed, twin boundaries between different domains in a single particle have been observed experimentally in monoclinic \ce{Li_{0.63}Ni_{1.02}O_2} by electron diffraction \cite{peres1999lithium} and also in \ce{Li_{0.5}CoO_2} by optical interferometric scattering microscopy \cite{merryweather2021operando}. Another interesting observation is the increased amount of tetrahedral \ce{Ni_{Li}} in the H2 phase. The cyan triangles in the center of three Li are Ni in the tetrahedral sites, and these Ni push the face-sharing Ni in the Ni layer to the adjacent Li layer forming Ni-Ni dumbbells. These dumbbells are only energetically favorable in the H2 phase. They disappear when charged to the H3 phase, but a small fraction could be kinetically trapped in the following discharge.  
 
\subsection{End of discharge}
Figure \ref{fig:discharge compare} (a) shows the discharge curves with different \ce{Ni_{Li}} from the simulations. The black curve with the lowest amount of \ce{Ni_{Li}} serves as the baseline with four profound plateaus A($\sim$4.1$\,$V), B($\sim$4.0$\,$V), C($\sim$3.7$\,$V), and D($\sim$3.5$\,$V). The voltage drops sharply below D. As y increases from 0.008 to 0.031, the shrink of plateau D results in an increase of the IRC, while the other plateaus remain almost the same. When y further increases to 0.063, the discharge starts with a slightly higher plateau A but drops more steeply after crossing the black curve at plateau C with the plateau D completely eliminated. The blue curve is more sloping compared to the other two, indicating the disturbance of Li-vacancy ordering by \ce{Ni_{Li}}. Figure \ref{fig:discharge compare} (b) shows the experimental results reproduced from Ref. \citen{phattharasupakun2021correlating}. The trends agree with our simulations: as y increases, D shrinks in capacity; A and B slightly shift up in voltage; and all the curves cross at C. These agreements confirm the fidelity of our model. 

% Temperature dependence
Figure \ref{fig:discharge compare} (c) and (d) compare the temperature dependence of the discharge at the minimal y. Temperature mainly affects plateau D as well in both the simulations and experiments: lower temperature shrinks plateaus D and thus increases the IRC. Raising the scan rate has a similar effect as lowering the temperature: both increase the Li concentration gradient within a particle. The blue dotted line at C/10 in Figure \ref{fig:discharge compare} (d) shows an even larger IRC compared to the red dashed line at C/20 at the same temperature. The temperature/scan-rate dependence further proves the kinetic root of the IRC.   

%\subsection{Atomistic origin of the kinetic hindrance}
\begin{figure}
    \includegraphics[width=0.9\columnwidth]{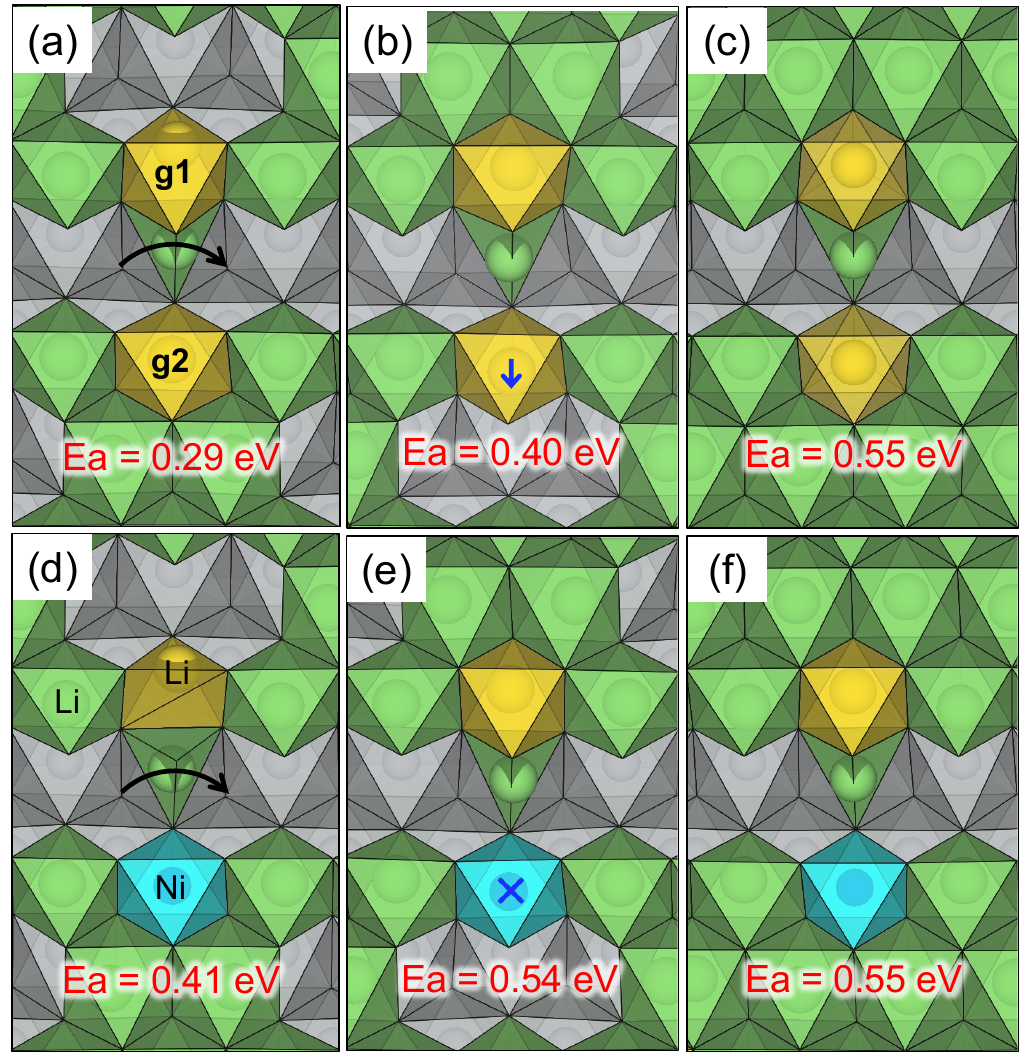}
	\caption{
	Comparison of Li hopping barriers with different occupancy of g1, g2 and their neighboring sites. The bottom row configurations have Ni in the g2 site. 
	\label{fig:barriers}
	}
\end{figure}
To understand the role of \ce{Ni_{Li}} and why it only manifests at plateau D, we need to first revisit the origin of the slow kinetics at the end of discharge. Li hops between octahedral sites through one of the two intermediate tetrahedral sites. The hopping barrier is determined by the energy difference between the tetrahedral and octahedral sites. The tetrahedral site energy strongly depends on the occupancy of its neighboring sites (i.e. gate sites).  Figure \ref{fig:barriers} (a) shows a Li going through a tetrahedral site along the black arrow. The top yellow gate site (g1) is its first nearest neighbor, and the bottom yellow gate site (g2) is its second nearest octahedral neighbor. In the conventional view, when the gate site g1 is occupied by another Li (the single-vacancy hop), the barrier is high due to the strong Coulomb repulsion; when g1 is not occupied (the di-vacancy hop), the  barrier is low \cite{van2001lithium}. However, it is necessary to go beyond this binary classification to a more continuous spectrum to fully explain the intricate Li transport phenomena. When Li content is over 50\%, the g1 site being empty becomes rare due to the Li-vacancy ordering. Figure \ref{fig:barriers} (a) shows an additional Li hopping between two Li rows, the ordering near 50\% Li content. Although the g1 site is occupied, the barrier is only 0.29$\,$eV, not much higher than the di-vacancy barrier of 0.27$\,$eV in a fully lithiated structure. Actually, Li at the g1 site shifts upwards significantly compared to its left and right neighbors to make way for the hopping Li. The energy penalty associated with the shift itself is small because the neighboring sites of g1 are empty. When the top neighboring sites of g1 are occupied, as shown in Figure \ref{fig:barriers} (b), the barrier increases by 0.11$\,$eV. When the four octahedral neighboring sites of g1 and g2 are fully occupied, as shown in Figure \ref{fig:barriers} (c), the barrier reaches 0.55$\,$eV, which resembles the single-vacancy barrier in a fully lithiated structure. Comparing (b) and (c), the occupation of the bottom neighboring sites of g2 raises the barrier by 0.15$\,$eV because the downward shift of the g2 Li is now suppressed. In summary, both g1 and g2 sites affect the barrier, and the gate Li atoms better maintain their positions when their backs are supported by other atoms. The growing overpotential at the end of discharge is caused by the increased occupation of the gate sites' neighbors, which prevents the gate from wide opening. The above picture is consistent with the di-vacancy argument but eliminates the need of analyzing the di-vacancy concentration or lifetime. The occupancy of the four neighboring sites of g1 and g2 is close to a uniform distribution. 

\ce{Ni_{Li}} is almost immobile compared to Li, like a rock in the river, which not only blocks Li hopping through it but also slows down Li moving around it. Ni at the g1 site strongly repulses the tetrahedral Li, but there are two tetrahedral sites between the two gate sites and Li can always hop through the one far from Ni. The following discussion will focus on Ni at the g2 site. Comparing Figure \ref{fig:barriers}(d)/(e) with (a)/(b), Ni substitution at the g2 site increases the barrier by 0.12$\,$eV and 0.14$\,$eV, respectively. More importantly, configuration (e) has a similar barrier as (c), but the former has no Li below g2 while the latter has two additional Li there. Roughly speaking, this single Ni has equivalent influence of three Li on adjacent diffusion paths. Ni at a gate site does not need the support from its neighbors to hold in place due to the stiffer Ni-O bonds. Therefore, the presence of \ce{Ni_{Li}} impedes Li transport without as many gate sites' neighbors being filled, which moves the sharp potential drop to lower Li content as y increases in Figure \ref{fig:discharge compare} (a). The stiffness of Ni at g2 is further confirmed in Figure \ref{fig:barriers} (f), where the barrier barely changes from (e) with two more Li added below the gate Ni. This Ni remains in the same place with or without the additional Li. The comparison between (f) and (c) implies that when the Li content is close to saturation, dilute \ce{Ni_{Li}} does not increase the Li hopping barrier further. Indeed, experimental diffusivity measurements show little impact of \ce{Ni_{Li}} below 3.6$\,$V \cite{phattharasupakun2021correlating}. When both the g1 and g2 sites are occupied by Ni, as shown in Figure S2 (e) in the SI, the barrier rises to 0.77$\,$eV, considerably higher than the values in Figure \ref{fig:barriers}. If the number of such Ni-Ni pairs increases, the Li diffusivity is expected to further decrease. 

Different from common belief, the repulsion between octahedral \ce{Ni_{Li}} and the nearest octahedral Li is not significantly stronger than the Li-Li repulsion. Figure S3 (a) and (b) in the SI show a Li hopping from the second nearest octahedral site to the nearest octahedral site of a \ce{Ni_{Li}} at two Li contents (both $>$ 50\%). The energy increase is 0.03$\,$eV and 0.06$\,$eV for the two configurations, respectively. Similar to the tetrahedral site case, the octahedral site repulsion also increases with the occupancy of far neighboring sites. The highest Ni-Li repulsion is 0.08$\,$eV, when there is only one vacant octahedral site next to the \ce{Ni_{Li}}. These numbers are relatively small and barely affect the equilibrium voltage profile, but they do add on top of the barrier for Li passing through. When \ce{Ni_{Li}} concentration is high, Li may become in contact with more than one \ce{Ni_{Li}} and feel stronger repulsion. Figure S3 (c) and (d) in the SI show Li hopping to a common neighbor of two \ce{Ni_{Li}}, and the energy increase is 0.07$\,$eV and 0.13$\,$eV, respectively. Similar to the influence on tetrahedral Li, \ce{Ni_{Li}} does not need support to hold in position and \ce{Ni_{Li}} pairs repulse Li more strongly. 

%\subsection{Asymmetry between charge and discharge}
\begin{figure}
    \includegraphics[width=0.9\columnwidth]{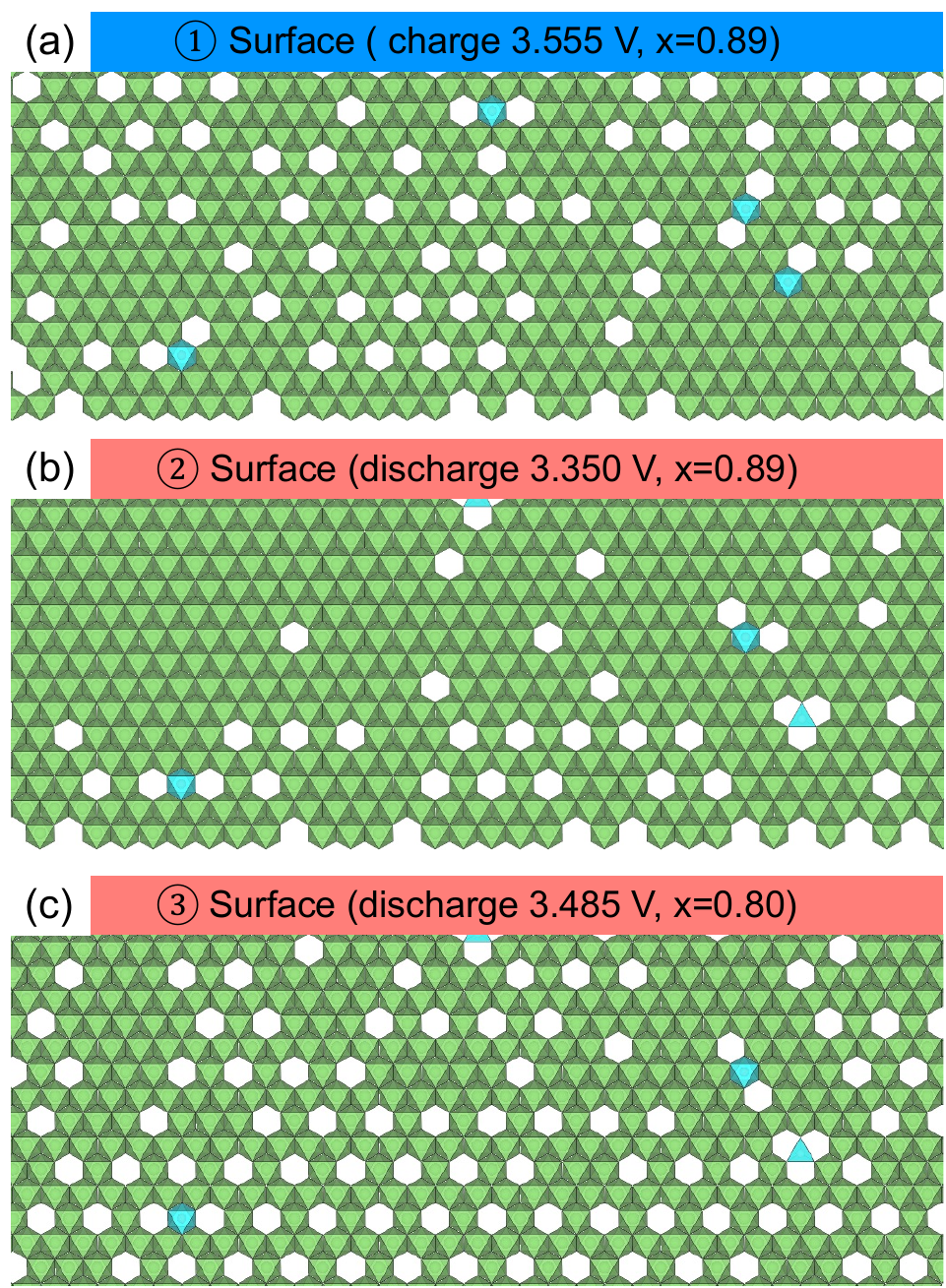}
	\caption{
	Snapshots of Li distribution near the surface at points \ding{172}, \ding{173} and \ding{174} in Figure \ref{fig:voltage_profiles}. 
	\label{fig:Li distributions}
	}
\end{figure}
One question remaining is why the overpotential of charge is lower than that of discharge at the same Li content near the fully lithiated state. This asymmetry arises from the dependence of Li hopping rate on the local occupancy combined with the vacancy concentration gradient formed during charge/discharge. Figure \ref{fig:Li distributions} (a) and (b) show the structure snapshots of a Li layer near the surface during charge and discharged with the same overall Li content, corresponding to the points \ding{172} and \ding{173} in Figure \ref{fig:voltage_profiles}. The charge configuration (a) has more vacancies near the surface, which lowers the diffusion barrier and enables fast Li supply from inside. These is a positive feedback during charge: the higher the overpotential, the higher the concentration gradient, and the higher the diffusivity as well. In contrast, the discharge configuration (b) has fewer vacancies near the surface spread far apart from each other. A hopping Li's gate sites are fully occupied, resulting in the highest diffusion barrier. This causes a dilemma during discharge: higher overpotential leads to a higher concentration gradient, but lower diffusivity. Galvanostatic current control would keep increasing the overpotential without reaching the desired current. Figure \ref{fig:Li distributions} (c) shows the structure snapshot at point \ding{174} in Figure \ref{fig:voltage_profiles}, before the sharp drop of discharge voltage. The near-surface Li vacancy concentration is clearly higher than that in Figure \ref{fig:Li distributions} (b),  confirming the connection between the growing overpotential and the near-surface Li saturation. The correlation between the self-limiting discharge kinetics and the concentration gradient within a single particle nickel-rich manganese cobalt oxide has been observed by operando optical microscopy recently \cite{XU20222535}.

%----------------------------------------------------------
\subsection{End of Charge}
Figure S4 in the SI shows the Li content (x) and voltage versus time for y=0.031. Delithiation of the last quarter of Li is extremely slow above 4.2$\,$V, with x dropping by 0.024 over \SI{300}{\micro\second}; while lithiation is much faster at around 4.1$\,$V, with x rising from 0 to 0.25 within \SI{300}{\micro\second}. This asymmetry at the end of charge has also been reported in a recent experimental study, where the charge voltage rises to 4.5$\,$V with a current of C/20 in GITT mode while the discharge voltage remains above 4.05$\,$V at the same composition in the first cycle \cite{bautista2023understanding}. Compared to the experimental equilibrium voltage of 4.15$\,$V, the charge overpotential is significantly higher than that of discharge at the same current. Our simulation suggests that this phenomenon is intrinsic to the pristine materials without degradation. 

\begin{figure}
    \includegraphics[width=\columnwidth]{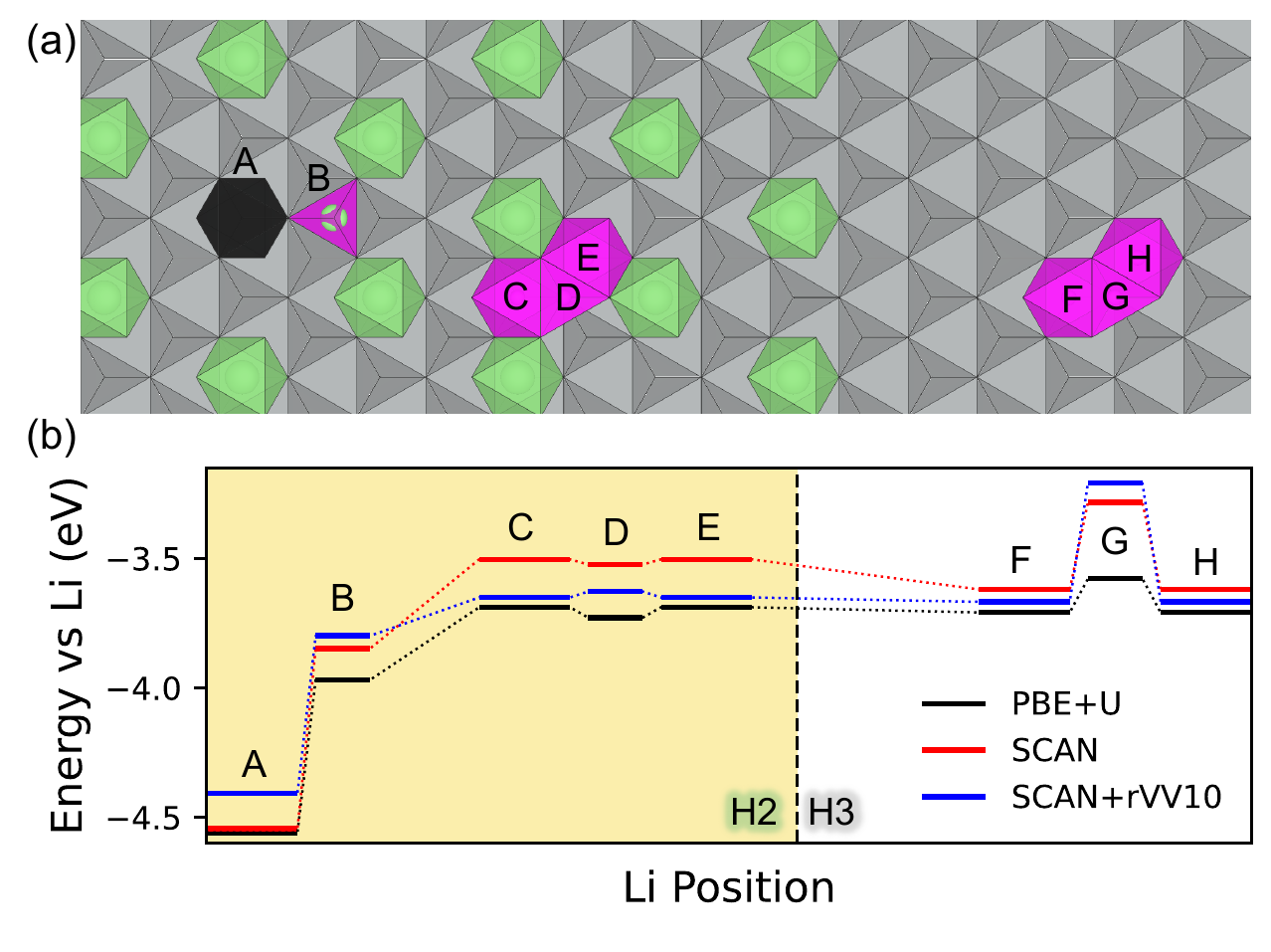}
	\caption{
	(a) a path of one Li moving from the H2 to H3 phase and (b) the corresponding energy landscape. The black site A represents the vacancy left; the magenta sites represent the interstitial sites the Li passing through. ACEFH are octahedral sites, and BDG are tetrahedral sites. Green octahedra are other Li ordered in the H2 phase.
	\label{fig:energy landscape}
	}
\end{figure}

\begin{figure*}
    \includegraphics[width=0.78\textwidth]{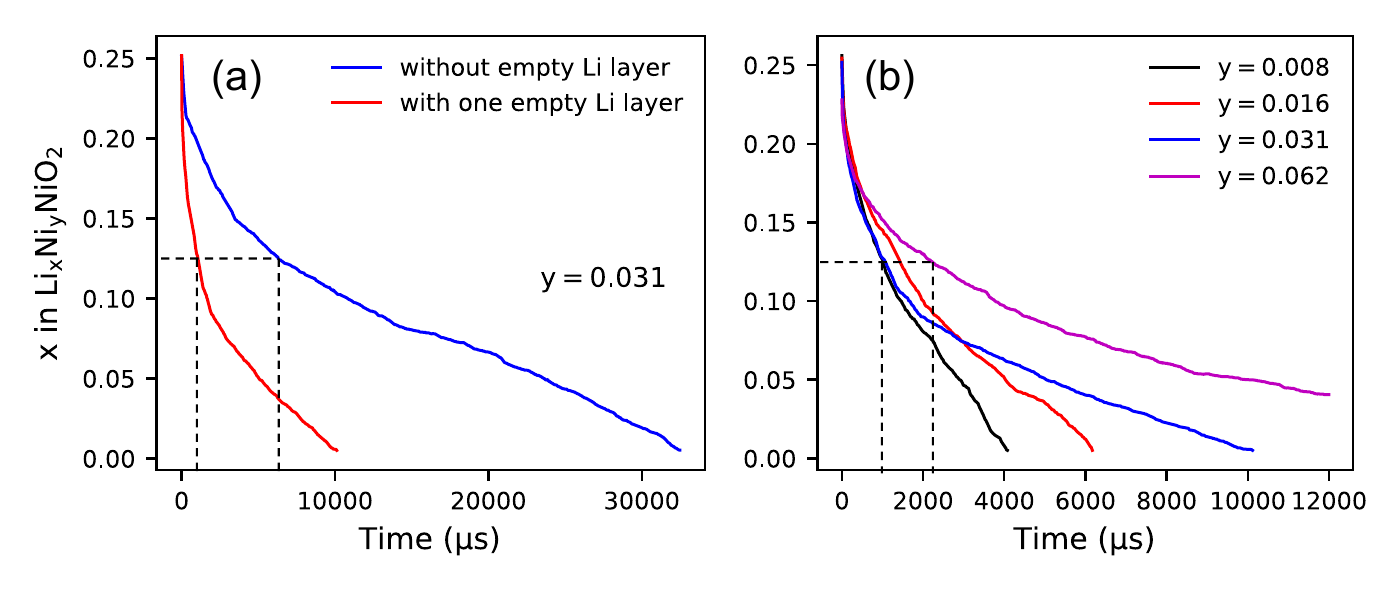}
	\caption{
	Li content as a function of time during charge at 4.4V. (a) effect of a preexisting empty layer; (b) effect of \ce{Ni_{Li}} with the preexisting empty layer. The dashed lines mark x=0.125, half capacity point of the plateau.
	\label{fig:charge compare}
	}
\end{figure*}

The slow kinetics at the end of charge has been attributed to low Li hopping rate in the H3 phase due to the shrinkage of layer spacing. The low probability of forming free Li, i.e. low carrier concentration, has not been considered before. Figure \ref{fig:energy landscape} shows the energy change of a Li moving from the H2 phase into the H3 phase, leaving a vacancy behind. When Li hops from A to B, a vacancy-interstitial pair is formed. Then the interstitial Li hops through CDE in the H2 phase and FGH in the H3 phase. The energy increase is 0.85$\,$eV between the two octahedral sites F and A, caused by breaking the effective Li-Li attractions, such as the linear Li-O-Ni-O-Li configuration across layers \cite{arroyo2002first}. This energy cost is higher than the hopping barrier in the H3 phase, i.e, the energy difference between G and F. Moreover, any Li in the H3 phase always has a chance to revert to the H2 phase. Prior to reaching the middle point between the H2/H3 interface and the particle surface, the likelihood of returning to the H2 phase is higher than escaping from the particle surface. Once reabsorbed in the H2 phase, forming another interstitial Li requires a long waiting period. As the H3 phase grows deeper into the particle, its growth rate decreases for the above reason. Since PBE+U overestimates the slab distance and  underestimates the Li hopping barrier, these energy levels are rechecked by SCAN and SCAN+rVV10. The latter two functionals predict the c lattice constant to be 13.87$\,$\AA and 13.39$\,$\AA, which are close to the experimental range (13.574-13.338$\,$\AA) \cite{li2018updating}. The Li hopping barrier in H3 increases to 0.33$\,$eV from SCAN and 0.46$\,$eV from SCAN+rVV10, respectively. %SCAN+rVV10 introduced a few more local minima with slightly different lattice parameters and the lowest ones are reported here. 
Our KMC simulations are based on PBE+U energies, with the hopping barrier in H3 measuring 0.13$\,$eV. The charging kinetics is already slow, and with a higher hopping barrier it would deteriorate further. Consequently, once the H3 phase forms a complete shell covering the surface during charge, additional Li extraction will encounter significant resistance. 

In contrast, during discharge Li can stick to the boundary of the H2 phase without hopping into the center of H3 phase by itself. This is reflected in the last snapshot in Figure \ref{fig:ordering}, where no isolated Li can be found in the center. Moreover, the energy levels of C and D are almost degenerate, and thus the interstitial Li hops extremely quickly in the H2 phase. The rate limiting step of discharge becomes Li injection, which is governed by the difference between Li chemical potential (-voltage) and the energy of C site, as shown in Eq.\ref{eq:BV}. The above provides another explanation for the asymmetry kinetics between charge and discharge in this regime despite the possible electronic conductivity difference \cite{bautista2023understanding}. %These possibilities can be checked experimentally. Adding more carbon black can eliminate the effect of electronic conductivity change; current as a function of voltage (Tafel plot) can be used to get more information. 

Breaking the effective Li-Li bonds is difficult, and thus the H3 phase is likely to grow along surfaces and defects where some of the Li-Li bonds are already broken. For example, the Li on the (003) surface have only half of the interlayer Li-Li bonds compared to those in the bulk; dislocations have been shown attract vacancies and could also interrupt the -Li-Li- chain \cite{sadowski2022planar}. Figure \ref{fig:charge compare} (a) shows the charging kinetics at 4.4$\,$V starting with and without an empty Li layer in the H2 phase. With this empty layer serving as the nucleation sites, the charging process is greatly accelerated. Similar behavior has been observed in \ce{LiFePO4}: the kinetics is improved by increasing the ratio of the (100) facet where the Li-Li attractions are terminated \cite{li2016100, xiao2018kinetic}. Another trend from Figure \ref{fig:charge compare} (a) is that the first half capacity takes much shorter time than the second half, which is the reason for the inaccessible Li at the end of charge. And these Li can be extracted under extremely low charging rates \cite{li2018updating, bautista2023understanding}. Figure \ref{fig:charge compare} (b) compares the kinetics with different amounts of \ce{Ni_{Li}}. Higher value of \ce{Ni_{Li}} extends the total charging time, mainly from the second half of the capacity. When a Ni atom substitutes a Li atom in the H2 phase, it attracts nearby Li atoms in a similar fashion as the Li. However, the Ni is not very mobile and will anchor some Li and slow down the phase boundary movement. 

%=========================================================================
\section{Discussion}
\begin{figure*}
    \includegraphics[width=0.8\textwidth]{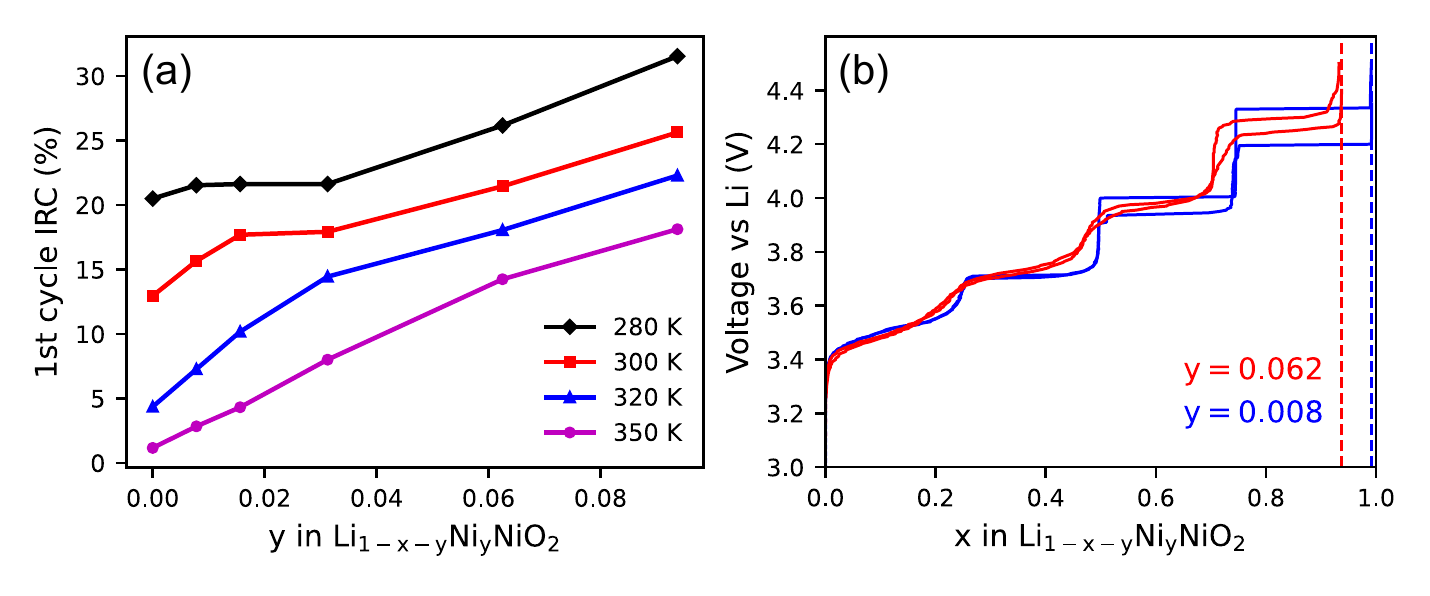}
	\caption{
	(a) IRC as a function of \ce{Ni_{Li}} (y) at different temperatures. (b) Equilibrium voltage profiles at two y values.  
	\label{fig:predictions}
	}
\end{figure*}

Figure \ref{fig:predictions} (a) summarizes the first cycle capacity loss as a function of \ce{Ni_{Li}} at different temperatures. At 350$\,$K where the concentration polarization is the lowest within the supercell, the capacity loss increases linearly with y. As the temperature is lowered and the concentration polarization increases, the curve shifts up non-uniformly. At 280$\,$K, the first part of the curve becomes almost flat for y$<$0.031. At this polarization level, the surface sites are saturated and the diffusion barrier is not affected by small amounts of \ce{Ni_{Li}}, as shown in Figure \ref{fig:barriers} (c) and (f). The linear relation has been reported \cite{phattharasupakun2021correlating}, where the tests were done at 303$\,$K with a C/20 rate; a nearly constant but higher capacity loss has also been reported for a similar primary particle size \cite{kurzhals2021linio2}, where the tests were done at 298 K with a C/10 rate. The seemingly contradictory results are probably caused by the difference in polarization within the particles. Both the lower temperature and higher scan rate lead to a higher Li concentration near the surface, and the diffusivity becomes insensitive to \ce{Ni_{Li}} near the fully lithiated state.

%\subsection{Thermodynamics vs. kinetics}
To better distinguish the thermodynamic and kinetic effects from \ce{Ni_{Li}}, every site is directly connected to the Li reservoir (grand canonical ensemble) to eliminate the effect of diffusion. The resulted voltage profiles are shown in Figure \ref{fig:predictions} (b). For x$<$0.2 the charge and discharge capacities are equal without any capacity loss. Also, the voltages near full lithiation barely change with y. This comparison further demonstrates that the IRC is caused by sluggish diffusion rather than thermodynamics. If the six neighboring sites of \ce{Ni_{Li}} were energetically unfavorable, the equilibrium voltage profile would have moved down significantly for x$\:<6\times $y. The transition from two-phase plateaus to more solid-solution-like slopes as y increases is still prominent in Figure \ref{fig:predictions} (b), indicating the thermodynamic root of such transition. Note that the small gaps between charge and discharge come from the phase-change hysteresis and the exact equilibrium voltages lie within the gaps. \ce{Ni_{Li}} facilitates phase nucleation and narrows the voltage gaps. 

%=========================================================================
\section{Conclusion}
We have devised a simulation framework that integrates non-equilibrium atomistic processes in battery electrodes during charge/discharge from first principles, which enables the kinetic competition between them without {\it a priori} assumptions. The simulation results on \ce{LiNiO_2} elucidate the origins of the first cycle IRC, the unattainable capacity at the end of charge, and the roles of \ce{Ni_{Li}} in these phenomena. The asymmetry between charge and discharge kinetics at both ends of the first cycle are captured as well. The sluggish discharge at high Li content is caused by the elevated diffusion barrier due to the increased occupancy of the gate Li's neighbors, while the sluggish charge at low Li content is caused by the frustration in H3 phase formation. \ce{Ni_{Li}} slows down both of them but for different reasons: during discharge, one \ce{Ni_{Li}} is equivalent to three Li in impeding Li diffusion; during charge, \ce{Ni_{Li}} traps Li to its second nearest octahedral sites. These findings offer new atomistic insights into the bottlenecks of Li transport in \ce{LiNiO_2} and high-Ni layered oxides.

%=========================================================================
%\section{Supporting Information}
%Supporting Information is available online.

%=========================================================================
\section{Acknowledgements}
P.X. and H.S.P acknowledge the support from the Natural Sciences and Engineering Research Council of Canada (NSERC) under the Discovery Grant RGPIN-2022-02969. The computational resources were provided by ACENET and the Digital Research Alliance of Canada. 
% -----------------------------------------------------------------
% Bibliography
\bibliography{dft,ce-kmc,LNO}
\end{document}

\begin{figure}
    \includegraphics[width=\columnwidth]{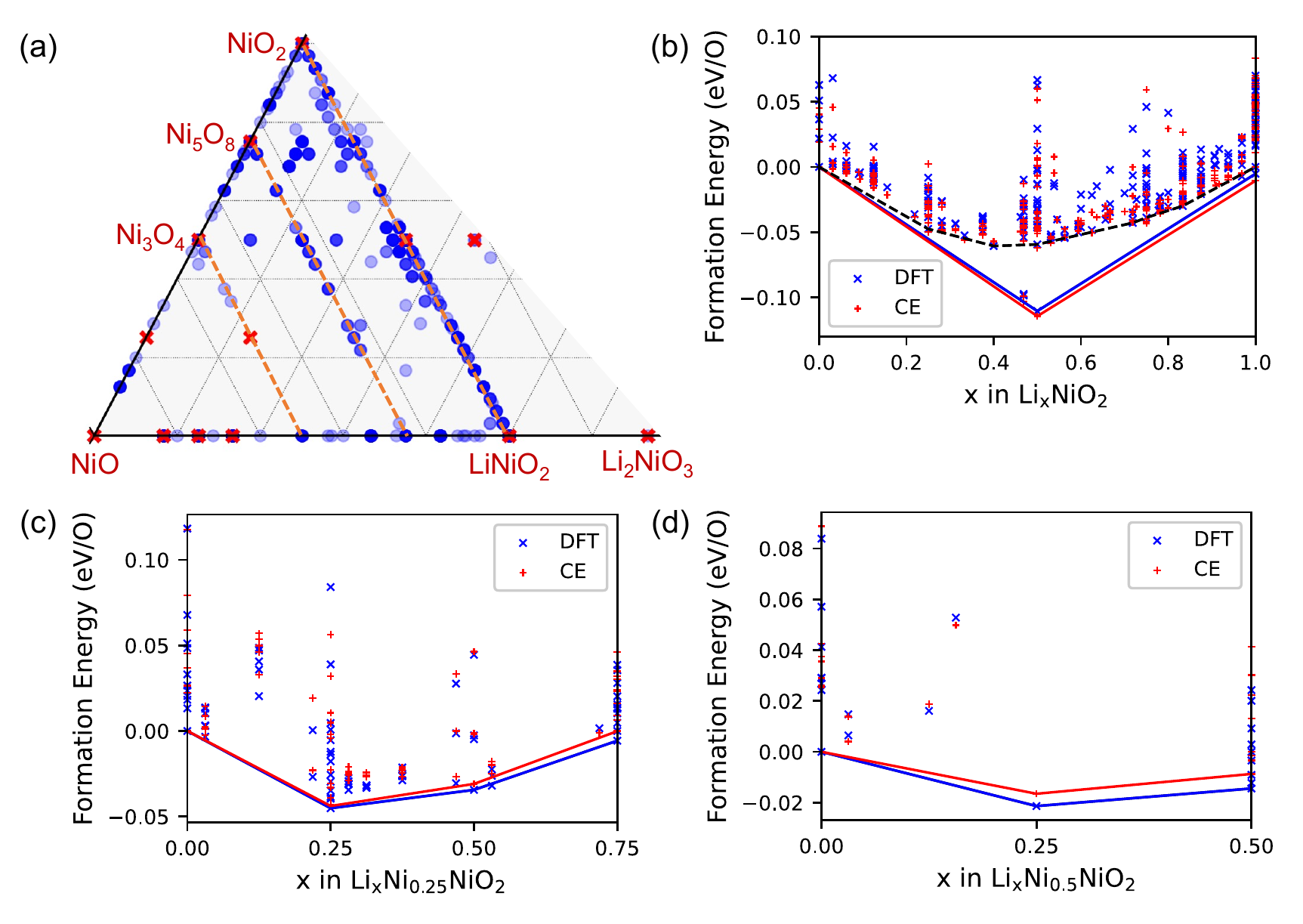}
	\caption{
	Cluster expansion (CE) training summary. (a) Composition of the training data; (b)-(d) energy comparison along the three dashed lines.
	\label{fig:hulls}
	}
\end{figure}

\begin{figure}
    \includegraphics[width=\columnwidth]{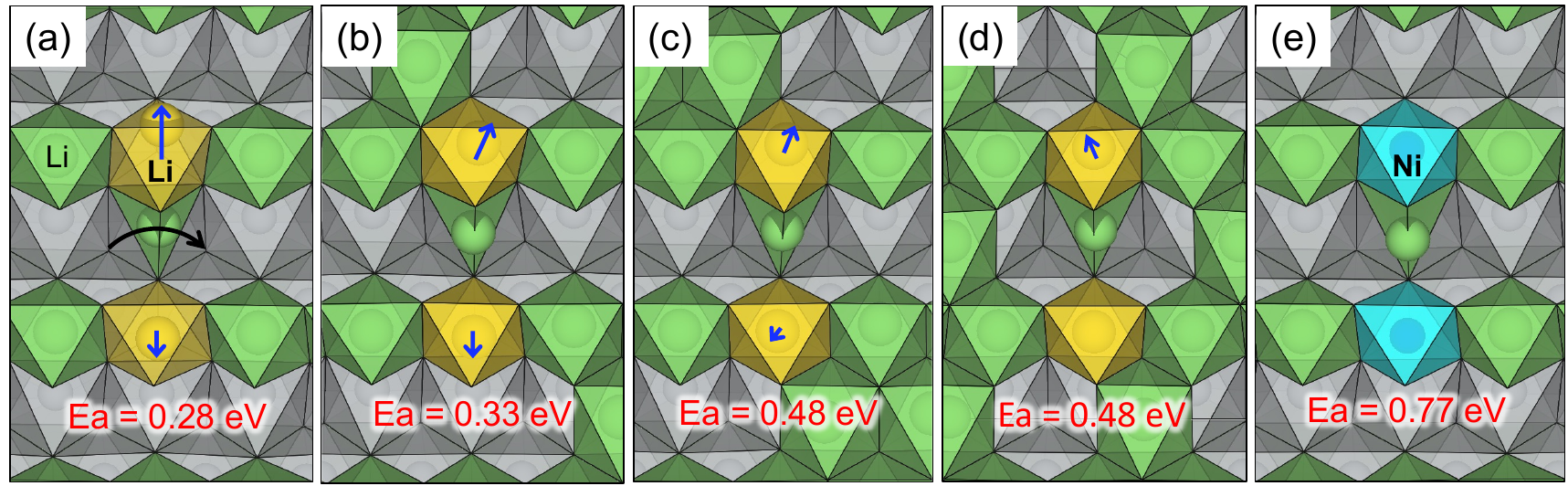}
	\caption{
	Additional Li hopping barriers. Two Ni occupying the two gate sites results in the highest barrier, even though their back neighbors are not occupied. 
	\label{fig:more barriers}
	}
\end{figure}

\begin{figure}
    \includegraphics[width=\columnwidth]{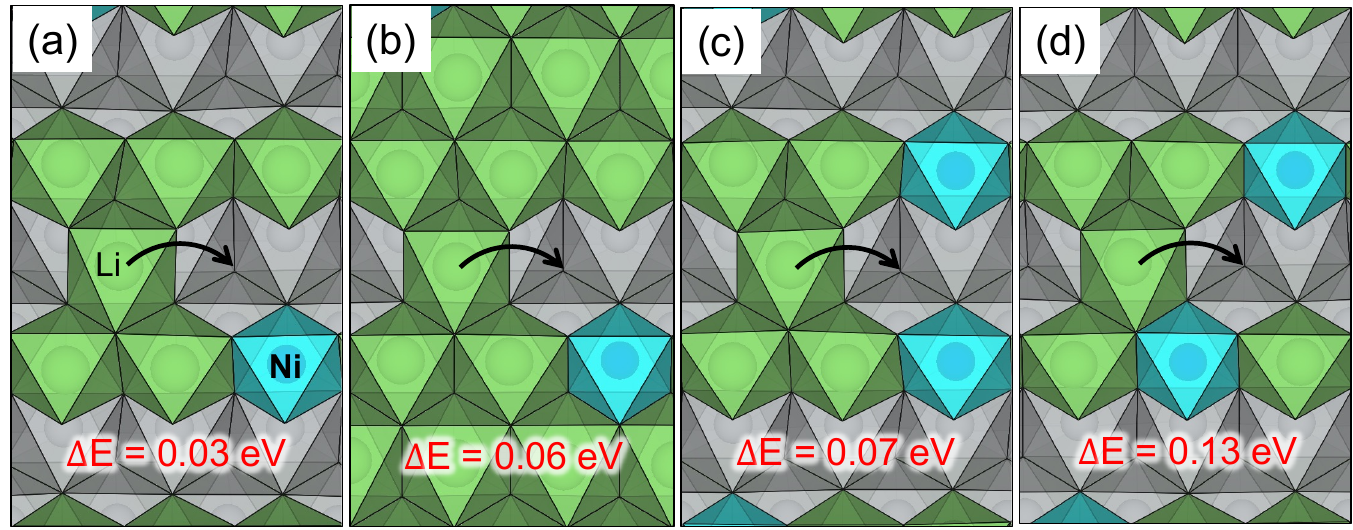}
	\caption{
	Repulsion between Li and Ni at neighboring octahedral sites at different Li occupancies. The repulsion increases with more Li or Ni nearby. 
	\label{fig:oct Li-Ni repulsion}
	}
\end{figure}

\begin{figure}
    \includegraphics[width=\columnwidth]{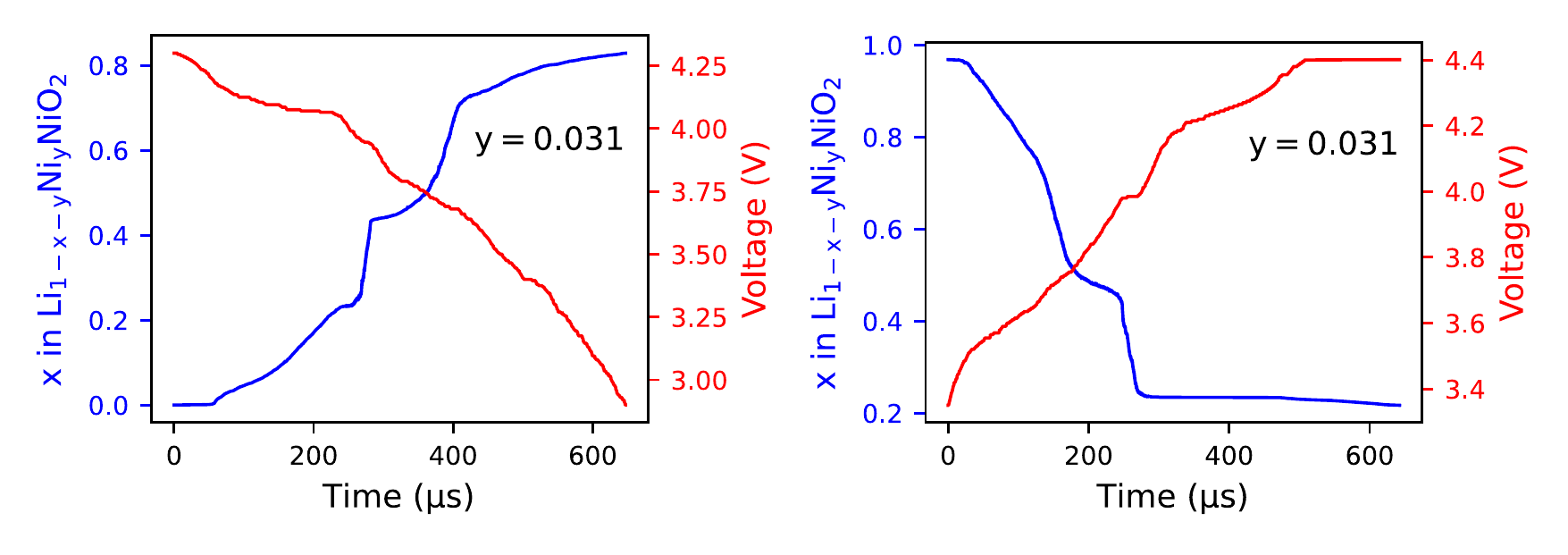}
	\caption{
	Overall Li content and voltage as a function of time during charge and discharge. The steeper change of x between 0.25 and 0.75 indicates extremely fast Li transport that exceeds the minimum current requirement under an overpotential around 5\,mV.  
	\label{fig:time evolution}
	}
\end{figure}